\documentclass[prl,twocolumn,floatfix,showpacs]{revtex4}

\usepackage{amsmath,amsthm,amsfonts,amssymb,bm}
\usepackage{epsfig}
\usepackage{times}
\usepackage{psfrag}

\newcommand{\mc}{ {m^{}_C} }
\newcommand{\me}{ {m^{}_E} }
\newcommand{\mr}{ {m^{}_R} }

\newcommand{\dind}{d_{\rm ind}}
\newcommand{\udes}{ L_S }

\newcommand{\ns}{{n^{}_S}}
\newcommand{\nc}{{n^{}_C}}

\newcommand{\nca}{{n^{}_{CA}}}
\newcommand{\nra}{{n^{}_{RA}}}

\newcommand{\normf}[1]{\|#1\|_{\rm fro}}

\newcommand{\Is}{I_S}
\newcommand{\Ic}{I_C}

\newcommand{\Ie}{I_E}

\newcommand{\favg}{f_{\rm avg}}

\newcommand{\adj}{{\dagger}}

\newcommand{\btab}{\begin{tabular}}
\newcommand{\etab}{\end{tabular}}

\newcommand{\rhos}{\rho^{}_{\rm S}}

\newcommand{\Rcal}{{\cal R}}

\newcommand{\trace}{{\bf Tr}}

\newcommand{\real}{{\rm Re}}

\newcommand{\gam}{\gamma}
\newcommand{\alf}{\alpha}

\newcommand{\bet}{\beta}
\renewcommand{\th}{\theta}

\newcommand{\rhoh}{\hat{\rho}}

\newcommand{\Ccal}{{\mathcal C}}

\newcommand{\Ecal}{{\mathcal E}}

\newcommand{\ie}{\emph{i.e.}}

\newcommand{\bquem}{\begin{quote}\begin{em}}
\newcommand{\equem}{\end{em}\end{quote}}

\newcommand{\blist}{\begin{description}}
\newcommand{\elist}{\end{description}}

\newcommand{\bquote}{\begin{quote}}
\newcommand{\equote}{\end{quote}}

\newcommand{\ben}{\begin{enumerate}}
\newcommand{\een}{\end{enumerate}}

\newcommand{\bit}{\begin{itemize}}
\newcommand{\eit}{\end{itemize}}

\newcommand{\bea}{\begin{array}}
\newcommand{\eea}{\end{array}}

\newcommand{\bds}{\begin{displaystyle}}
\newcommand{\eds}{\end{displaystyle}}

\newcommand{\refeq}[1]{(\ref{eq:#1})}

\makeatletter
\def\beq{\@ifnextchar 
[{\@tempswatrue\@beq}{\@tempswafalse\@beq[]}}
\def\@beq[#1]{\begin{equation}\edef\@tmparg{#1}\ifx\@tmparg\@e
mpty \else
	\label{#1}\fi}
\newcommand{\eeq}{\end{equation}}
\makeatother 
\newcommand{\beqaa}{\begin{eqnarray*}}
\newcommand{\eeqaa}{\end{eqnarray*}}

\newcommand{\beqa}{\begin{eqnarray}}
\newcommand{\eeqa}{\end{eqnarray}}

\newcommand{\bc}{\begin{center}}
\newcommand{\ec}{\end{center}}

\usepackage[dvips]{color}

\begin{document}
\title{ Robust Quantum Error Correction via Convex Optimization }
\author{Robert L. Kosut$^{(1)}$, Alireza Shabani$^{(2)}$, Daniel A. Lidar$^{(2,3,4)}$}
\affiliation{$^{(1)}$SC Solutions, Inc., 1261 Oakmead Parkway, Sunnyvale, CA 94085}
\affiliation{Departments of $^{(2)}$Electrical Engineering, 
$^{(3)}$Chemistry, and $^{(4)}$Physics, University of Southern California, 
Los Angeles, CA 90089}

\begin{abstract}
We present a semidefinite program optimization approach to quantum error
correction that yields codes and recovery procedures that are robust
against significant variations in the noise channel. Our approach allows us to optimize
the encoding, recovery, or both, and is amenable to approximations
that significantly improve computational cost while retaining
fidelity. We illustrate our theory numerically for
optimized 5-qubit codes, using the standard [5,1,3] code as a
benchmark. Our optimized encoding and recovery yields fidelities that
are uniformly higher by 1-2 orders of magnitude against random unitary weight-2 errors
compared to the [5,1,3] code with standard recovery. We observe
similar improvement for a 4-qubit decoherence-free subspace code
\end{abstract}


\maketitle

%%%%%%%%%%%%%%%%%%%%%%%%%%%%%%%%%

\textit{Introduction}.--- Quantum error correction (QEC) is essential for
the scale-up of quantum information devices. A theory of quantum error
correcting codes has been developed, in analogy to classical coding
for noisy channels
\cite{Shor:95Gott:96Steane:96,Laflamme:96,KnillL:97,NielsenC:00}.
This theory allows one to find perfect-fidelity encoding and recovery procedures for a
wide class of noise channels. While it
has led to many breakthroughs, this
approach has two important disadvantages: (1) Robustness: Perfect QEC schemes, as well those produced by optimization tuned to
specific errors, are often not robust to even small changes in the
noise channel. (2) Cost: The encoding and recovery effort in perfect QEC typically grows
exponentially with the number of errors in the noise channel. Here we
present an optimization approach to QEC that addresses both these
problems. Regarding robustness, we develop an approach that incorporates specific models of
noise channel uncertainty, resulting in highly robust error
correction. Thus entire classes of noise channels
which do not satisfy the standard assumptions for perfect correction
\cite{Shor:95Gott:96Steane:96,Laflamme:96,KnillL:97,NielsenC:00}, can
be tailored with optimized encoding and/or recovery.
Regarding cost, if the resulting robust fidelity levels are sufficiently high, then
{\em no further increases in codespace dimension and/or levels of
concatenation are necessary}. This assessment, which is critical to
any specific implementation, is not knowable without performing the
robust optimization. 

\textit{Relation to prior work}.--- 
An optimization approach to QEC was reported in a number of recent papers
\cite{ReimpellW:05,YamamotoHT:05FletcherSW:06ReimpellWA:06,KosutL:06}.
In these works error correction
design was posed as an optimization problem
to {\em directly} maximize fidelity, with the design variables being
the process matrices associated with the encoding and/or recovery
channels. Here we present an {\em indirect} approach to fidelity
maximization based on minimizing the error between the actual channel
and the desired channel.  Both the direct and indirect approaches lead
naturally to bi-convex optimization problems, specifically, two
semidefinite programs (SDPs) \cite{BoydV:04} which can be iterated
between recovery and encoding. For a given encoding the problem is
convex in the recovery. For a given recovery, the problem is convex in
the encoding. However, the indirect approach is in general
more efficient computationaly then the direct approach and, as we will
show, has the added advantage of incorporating an approximation method
which provides a considerable reduction in computational cost with
only a minimal fidelity loss.

\textit{Organization}.--- After reviewing the standard error correction
model and defining performance measures, we state the direct and indirect
fidelity optimization problems, then show how to add robustness
measures, and describe methods to solve the (robust) indirect problem.
We present a range of examples illustrating the robustness of our
codes against increasingly more challenging noise channels, and
conclude with a discussion and summary of computational cost.

%%%%%%%%%%%%%%%%%%%%%

\textit{Noise and error correction model}.--- Subject to standard
assumptions, the dynamics of any open quantum system can be described
in terms of a completely-positive (CP) map: $\rho_S \to \sum_i A_i
\rho_S A_i^\dagger$, a result known as the Kraus Operator Sum
Representation (OSR) \cite{NielsenC:00}. Here $\rho_S$ is the initial
system density matrix and the $A_i$ are called operation elements, and
satisfy $\sum_i A_i^\dagger A_i = I_S$ (identity). The standard error
correction procedure involves CP encoding ($\Ccal$), error ($\Ecal$),
and recovery ($\Rcal$) maps (or channels): $\rho _{S}\overset{\Ccal
}{\to}\rho _{C}\overset{\Ecal}{\to }\sigma _{C}\overset{\Rcal}{\to
}\hat{\rho}_{S}$, i.e., using the OSR: $\rhoh_S = \sum_{r,e,c} (R_r
E_e C_c) \rhos (R_r E_e C_c)^\dag$ (see Ref.~\cite{SL:07} for a
relaxation of the CP map condition).  The encoding $\{C_c\}_{c=1}^\mc$
and recovery $\{R_r\}_{r=1}^\mr$ operation elements are rectangular
matrices, respectively $\nc\times\ns$ and $\ns\times\nc$, since they
map between the system Hilbert space (of dimension $n_S$) and the
system+ancillae Hilbert space (of dimension $n_C$). The error
$\{E_e\}_{e=1}^\me$ operation elements are square $(\nc\times\nc)$
matrices, and represent the effects of decoherence and noise.  As in
\cite{KnillL:97}, we will restrict attention to {\em unitary
encoding}: $\Ccal$ has only a single OSR element, the $\nc\times\ns$
{\em encoding matrix} $C$, whose $\ns$ columns are the orthonormal
\emph{codewords} ($C^\dag C=\Is$) with $\nc=\ns\nca$, $\nca$ being the
dimension of the encoding ancilla space.  It follows from $\rho_S =
{\rm Tr}_{CA,RA}[U_R (|0\rangle\langle 0|_{RA} \otimes \sigma_C ) U_R^\dagger ]$,
  where $U_R$ is the unitary recovery acting on the noisy encoded state $\sigma_C$ and
  the recovery ancillae $|0\rangle_{RA}$,
  that $\mr=\nca\nra$,
with $\nra$ the dimension of the recovery ancillae space.
%%%%%%%%%%%%%%%%%%%%%%%%%%%

{\it Performance measures}.--- Our error correction objective
is to design the encoding $\Ccal$ and recovery $\Rcal$
so that, for a given $\Ecal$, the map $\rho_S \to \hat{\rho}_S$ is as
close as possible to a desired $\ns\times\ns$ unitary $\udes$.  A
common measure of performance is the \emph{average fidelity} between
the channel $\Rcal\Ecal\Ccal$ and the ideal $\udes$: $\favg = \frac{1}{n_s^2} \sum_{r,e}\ 
|\trace\ \udes^\dag R_r E_e C |^2$.  
$\favg=1$ if
and only if there are constants $\alf_{re}$ such that
\cite{KnillL:97,NielsenC:00}: $R_r E_e C = \alf_{re} \udes,\ \sum_{r,e} |\alf_{re}|^2 = 1$.
This suggests the \emph{indirect} measure of fidelity, the
``distance-like'' error (using the Frobenius norm, $\normf{X}^2=\trace\
X^\dag X$),
\beq[eq:dind]
\bea{rcl}
\dind &=& \sum_{r,e}\ \normf{ R_r E_e C_c - \alf_{re} \udes }^2 
\\
&=& \normf{ R E(\Ie\otimes C)-\alf\otimes \udes }^2
\eea
\eeq
with $\alf$ the $\nca\nra \times {m^{}_E} $ matrix with elements
$\alf_{re}$, $E$ the $\nc\times\nc {m^{}_E} $ rectangular {\em error
system matrix} $E=[E_1\ \cdots\ E_ {m_E} ], $ and $R$ the
$\nc\nra\times\nc$ {\em recovery matrix} obtained by stacking the
$\nca\nra$ matrices $R_r$. Hence, we have $\normf{\alf}^2=\trace\
\alf^\dag\alf=\sum_{r,e}|\alf_{re}|^2=1$, and $R^\dag R=\sum_r\
R_r^\dag R_r = \Ic $. Note that $U_R = [R|W]$, where $U_R$ is the
unitary recovery and $W$ is $\nc\nra\times(\nc\nra-\nc)$.

As $\favg$ and $\dind$ are explicitly dependent on the channel
elements, they are convenient for optimization. Consider then the
following optimization problems.
\beq[eq:favg opt]
\bea{c} 
\mbox{\em Direct Fidelity Maximization} 
\\
\bea{ll} 
\mbox{maximize} 
& 
\favg=\frac{1}{n_s^2}
\sum_{r,e}\ | \trace\ \udes^\dag R_r E_e C |^2 
\\
\mbox{subject to} 
& 
R^\dag R = \Ic, \;\; C^\dag C = \Is
\eea
\eea
\eeq
\vspace{-3ex}
\begin{eqnarray}
\label{eq:dind opt}
\bea{c} 
\mbox{\em Indirect Fidelity Maximization} \\
\bea{ll} 
\mbox{minimize} 
& 
\dind=\normf{ R E(\Ie\otimes C)-\alf\otimes \udes }^2 
\\
\mbox{subject to} 
& 
R^\dag R = \Ic, \;\; C^\dag C = \Is, 
\;\;
\normf{\alf}^2=1 
\eea
\eea
\end{eqnarray}
Both are non-convex optimization problems for which local solutions
can be found from a bi-convex iteration.  The direct problem was
addressed in \cite{ReimpellW:05,YamamotoHT:05FletcherSW:06ReimpellWA:06,KosutL:06}.
We now discuss methods to obtain local solutions to the indirect
problem.

%%%%%%%%%%%%%%%%%%%%%%%%%%%%%%%%%%%%%%%
{\it Robust error correction}.--- A major limitation of the standard
procedure of modeling the error channel as fixed, \ie, in terms of
given operation elements $\{E_e\}$, is that this does not account for
uncertainty in knowledge of the channel, and in most cases will hence
be too conservative. For example: different runs of a tomography
experiment can yield different error channels
$\{\Ecal_\beta\}_{\bet=1}^\ell$; an OSR model $\Ecal(p)$ could depend
on an uncertain parameter $p$; a physical model of the error channel
might be generated by a system-bath Hamiltonian $H(\th )$ dependent
upon an uncertain set of parameters $\th $. Whatever the source or
sources, not accounting for model uncertainties typically leads to
non-robust error correction, in the sense that a small change in the
error model can lead to poor performance of the error correction
procedure. One way to account for uncertainties in terms of an OSR is
to take a sample from the set, say, $\{\Ecal(p_\bet)\}_{\bet=1}^\ell$
or $\{H(\th_\bet)\}_{\bet=1}^\ell$.  In the latter case, tracing out
the bath states will result in a \emph{set} of error system matrices
$\{E_\bet \}_{\bet=1}^\ell $.  To handle this, the objectives in
\refeq{favg opt}-\refeq{dind opt} need to be modified. Two
possibilities are the \emph{ worst-case} and \emph{average-case}. For
the worst-case, these objectives can be replaced by optimizing over
all $E_\bet$.  In the average-case, the objectives can be equivalently
expressed in the same form but with $\favg$ and $\dind$ replaced by
their average over all $E_\bet$. This is equivalent to replacing the
error system matrix elements by $E_e \mapsto E_{\bet,e}/\sqrt{\ell }$.

%%%%%%%%%%%%%%%%%%%%%%%%%%%%

{\it Indirect fidelity maximization}.--- Using the constraints in
\refeq{dind opt} gives the distance measure \refeq{dind} as,
\begin{eqnarray}
&&\dind
= \normf{ R E (\Ie \otimes C) - \alf \otimes
\udes}^2 \label{eq:obj01} \\
&&=
\ns +\trace\ E(\Ie\otimes C C^\adj)E^\adj
-2\real\ \trace\ RE(\alf^\adj\otimes C \udes^\adj) \notag
\end{eqnarray}
Since only the last term depends on $R$, minimizing $\dind$ over $R$
is equivalent to maximizing the last term over $R$. A singular value
decomposition of the $\nc\times\nc\nra$ matrix $E(\alf^\adj\otimes C
\udes^\adj)$ immediately yields,
\beq[eq:maxvr01]
\max_{R^\adj R=\Ic}\ \real\ 
\trace\ RE(\alf^\adj\otimes C \udes^\adj) = \trace\ \sqrt{ E(\gam\otimes C C^
\adj)E^\adj}
\eeq
with the ${m_E} \times {m_E} $ matrix $\gam=\alf^\dag\alf$. The $\nc
\nra\times\nc$ optimizing recovery matrix $R$ is given by:
\beq[eq:ropt ind]
R = \left[v_1\ \cdots\ v_{\nc}\right]\ [u_1\ \cdots\ u_\nc]^\dag ,
\eeq
where the $v_i$ and $u_i$ are, respectively, the right and left
singular vectors of the matrix $E(\alf^\adj\otimes C \udes^\adj)$.
Given $(C,\gam)$, and the fact that $\alf$ need only be chosen so that
$\alf^\dag\alf=\gam$, the following choice for $\alf$ achieves $\gam$.
\begin{eqnarray}
\bea{rcl}
\nca &\geq& \me 
\\ 
\nra &=& 1 
\eea
&\Rightarrow&
\bea{l}
\alf=\left[\bea{c} \sqrt{\gam} \\ 0_{\nca-\me\times\me}\eea\right]
\\
\mbox{$R$ is $\ns\nca\times\ns\nca$ (unitary)}
\eea
\label{eq:alfopt a}
\\
\bea{rcl}
\nca &<& \me 
\\ 
\nra\nca &=& \me
\eea
&\Rightarrow&
\bea{l}
\alf=\sqrt{\gam}
\\
\mbox{$R$ is $\ns\me\times\ns\nca$ (tall)}
\eea
\label{eq:alfopt b}
\end{eqnarray}
Result \refeq{alfopt a} implies that $R$ is unitary when the number of
encoding ancillas, $\nca$, is chosen large enough that no recovery
ancillas are needed, \ie, $\nra=1$, and $U_R=R$.
Related results about unitarily recoverable codes
were obtained in \cite{KribsS:06}.  When there are insufficient
encoding ancilla, \ie, $\nca < \me$, \refeq{alfopt b} reveals that
additional recovery ancilla are needed so that $\me=\nca\nra$.  The
result in \refeq{alfopt a}-\refeq{alfopt b} does not change if $\alf$
multiplied by a unitary. 
This unitary freedom is exactly the unitary freedom in choosing the
OSR operators \cite{NielsenC:00}. Note also that the relation
$\nra\nca=\me$ may require that $R$ is padded with zeroes.

%%%%%%%%%%%%%%%%%%%%%%%%%%%%%

{\it Optimal recovery}.--- Given an encoding $C$, an optimal recovery
$R$ can be obtained in two steps. 
\bquote
\vspace{-2ex}
Step 1: solve for $\gam$ which maximizes \refeq{maxvr01}, that is,
\beq[eq:gamopt] 
\bea{ll} 
\mbox{maximize} & \trace\ \sqrt{E(\gam\otimes C C^\adj)E^\adj} 
\\
\mbox{subject to} & \gam \geq 0, \;\; \trace\ \gam=1 
\eea
\eeq
\vspace{-2ex}
Step 2: obtain $R$ from $\gam$ via \refeq{ropt ind}-\refeq{alfopt b}.
\equote 
Since the negative of the objective function in Step 1 satisfies the
second order condition for convexity, and the constraint is a convex
set in $\gam$, it follows that \refeq{gamopt} is a convex optimization
problem \cite{BoydV:04}.

\iffalse
Since $R$ is $\nra\nc\times\nc$, it becomes square, and hence unitary, when
the number of recovery ancilla $\log_2\nra=0$. It follows from 
\refeq{alfopt
a}-\refeq{alfopt b} that this is the case when the dimension of the encoding
ancilla, $\nca$, is no smaller than ${m_E} $, the number of error
operators. 
\fi

%%%%%%%%%%%%%%%%%%%%%%%%%%

{\it Approximation to optimal recovery}.--- If the errors were random
unitaries, \ie, $E_i = \sqrt{p_i} U_i$, where $\{ p_i \}$ are
probabilities and $\{ U_i \}$ are unitaries, then the diagonal
elements of the matrix $\gam$ would correspond to the probability of
the associated error \cite{ZanardiLidar:04}. Generalizing to arbitrary
channels, we consider the {\em approximation} of setting $\gam$ equal
to the diagonal matrix with diagonal elements,
\beq[eq:gam approx]
\gam_{ii} = \normf{E_i}^2/\nc
\eeq
each being the average sum-square of the singular values of
$E_i$. Since $\Ecal$ is trace-preserving, $\sum_i\gam_{ii}=1$ as
required.  Using this approximation in \refeq{alfopt a}-\refeq{alfopt
b} to calculate $R$ directly is obviously very efficient, especially
for large dimensions, where by comparison solving \refeq{gamopt} can
be computationally expensive even when $\gam$ is constrained to be
diagonal.

%%%%%%%%%%%%%%%%%%%%%%%%

{\it Optimal encoding}.--- Given $(R, \alf)$, an optimal encoding can
be found by solving \refeq{dind opt} for $C$. Replacing the non-convex
equality constraint, $C^\dag C=\Is$, by the convex inequality $C^\dag
C\leq \Is$ leads to the \emph{relaxed} convex optimization problem,
\beq[eq:coptrlx] 
\bea{ll} 
\mbox{minimize} 
& 
\dind=\normf{R E (\Ie \otimes C) - \alf \otimes \udes}^2 
\\
\mbox{subject to} & C^\adj C \leq \Is
\eea
\eeq
By replacing the singular values of the optimal relaxed solution to
\refeq{coptrlx} with ones, we obtain a nearby encoding which satisfies
$C^\dag C=\Is$.
%\cite{GolubV:83}.

%%%%%%%%%%%%%%%%%%%%%%%%%%%
{\it Examples}.--- We now apply the methods developed above to the
goal of preserving a single qubit ($\ns=2$) using $4$ and $5$-qubit
codespaces. We consider noise channels with weight-2 and 3 errors, and
compare optimal encoding and recovery to the performance of the
$[5,1,3]$ code, which is perfect against arbitrary weight-1 errors
\cite{Laflamme:96}, and to the 4-qubit DFS code (denoted ``DFS-4'')
which is perfect against collective errors \cite{Zanardi:97c}. We
consider an independent errors model, where an error on $t>0$ qubits
has probability 
$P(t)=p^t(1-p)^{\nc-t}/Z$, $Z=\sum_{t=0}^w P(t)$, where $w$ is the weight.
For 5 qubits with weight-2 errors there are
$\me=\sum_{t=0}^2 \binom{5}{t}=16$ OSR error system elements.
For weight-3 errors $\me=26$.

\begin{figure}[tbp]
\btab{c}
\epsfig{
file=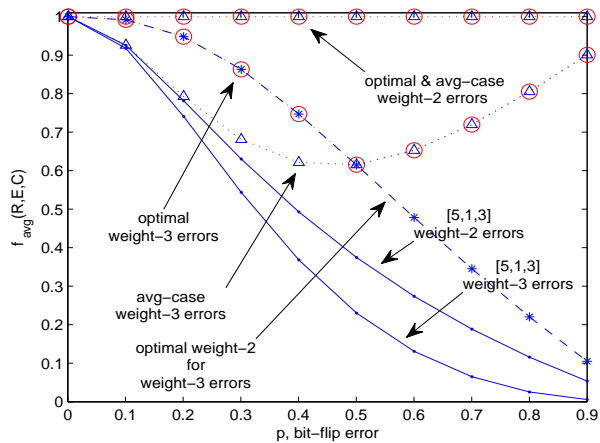,
width=3.5in,height=2.5in} 
\etab
\vspace{-3ex}
\caption{Weight-2 \& 3 bit-flip error for the $[5,1,3]$ code, with
  different recoveries.}
\label{fig:5qu_bf}
\end{figure}

%%%%%%%%%%%%%%%%%%%%%%%
{\it Optimized recovery for weight-2,3 bit-flip errors}.---
Fig.~\ref{fig:5qu_bf} shows $\favg(R,E,C)$ vs. bit-flip probability
$p$ with weight-2 and weight-3 errors. In this figure the code is
always the standard $[5,1,3]$ code and we compare different recoveries.
The solid lines show the $[5,1,3]$ code performance with standard
recovery for arbitrary weight-2 or weight-3 errors, which as expected,
is not good. The curve with circle markers labeled ``optimal weight-3
errors'' is for an optimal recovery by solving \refeq{gamopt} for the
optimal $\gam$ using the standard $[5,1,3]$ encoding and then finding $R$ from
\refeq{alfopt a}-\refeq{alfopt b}. The improved performance over the
solid lines is entirely due to the optimized recovery. Iteration between encoding and recovery
[Eqs.~\refeq{alfopt a}-\refeq{coptrlx}] changed neither the optimized
recovery nor the $[5,1,3]$
encoding. We also see that the optimal performance decreases until
$p=0.5$ and then increases. The average-case recovery
indicated by the dotted curve with triangle markers falls below the
optimal for $p<0.5$ and then is identical with the optimal for $p\geq
0.5$. 
For weight-2 errors, perfect fidelity over the entire $p$-range is
achieved by the optimal and average-case recovery (labeled
``optimal \& avg-case, weight-2 errors'') which are found by
solving either \refeq{gamopt} for the optimal $\gam$ or using the
approximation \refeq{gam approx} and then finding $R$ from
\refeq{alfopt a}-\refeq{alfopt b}. The perfect performance is easily
understood from the fact that the $[5,1,3]$ code also perfectly corrects
weight-2 bit-flip errors. Our optimization finds the corresponding
ideal recovery. When this recovery is applied to the weight-3
errors we get the dashed curve with star markers labeled ``optimal
weight-2 for weight-3 errors''.  This curve follows the weight-3 optimal recovery for
$p<0.5$ and diverges thereafter, a phenomenon similar to what was
reported for amplitude-damping errors in \cite{ReimpellW:05}.
In Fig.~\ref{fig:5qu_bf}, in all weight-2 cases the recovery matrix $R$ was a
$2^5\times 2^5$ unitary, or easily reducible to that via a singular
value decomposition, i.e., no recovery ancillae were needed [recall
  Eq.~\refeq{alfopt a}]. For weight-3 errors, $\me=26$ requires two
recovery ancillas, as per Eq.~\refeq{alfopt b} with $\nra\nca=\me$.

\begin{figure}[tbp]
\btab{c} 
\epsfig{
file=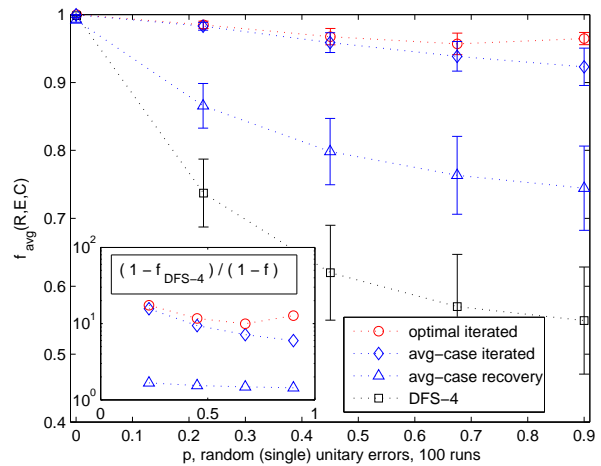,
width=3.5in} 
\etab
\vspace{-3ex}
\caption{Weight-2 single-random unitary errors for 4-qubit codes.}
\label{fig:5qu_ue_standard}
\end{figure}

\begin{figure}[tbp]
\btab{c} 
\epsfig{
file=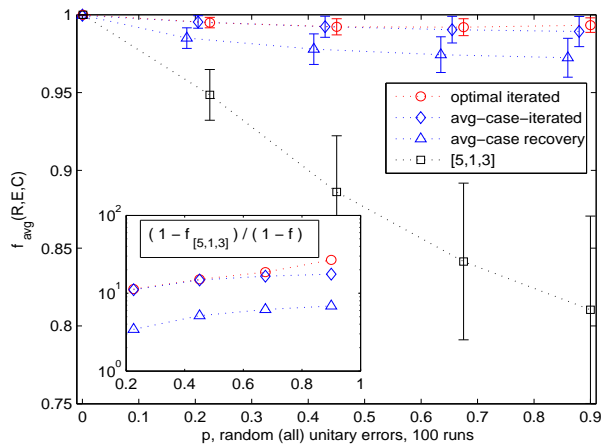,
width=3.5in,height=2.5in} 
\etab
\vspace{-3ex}
\caption{Weight-2 all-random unitary error for 5-qubit codes.}
\label{fig:5qu_ue_allrand}
\end{figure}

%%%%%%%%%%%%%%%%%%%%%%%
{\it Weight-2 random unitary errors}.---
Here, with the
exception of results marked by ``iterated'', we always optimized
recovery for the DFS-4 or the $[5,1,3]$ code. In the ``iterated'' case we start
from the DFS-4 or the $[5,1,3]$ code and optimize the encoding as
well. We examined two cases: (1) for the DFS-4 code, a single unitary is
randomly chosen and used for every one of the 15 errors, thus yielding
collective weight-2 errors; (2) for the
$[5,1,3]$ code, 15 unitaries are independently randomly
selected. While the choice of the DFS-4 code is somewhat arbitrary, we
picked this code for case (1) since the errors in this case are
similar to the collective decoherence model \cite{Zanardi:97c}.
Figure \ref{fig:5qu_ue_standard} shows the mean fidelity and standard
deviation (error bars) from 100 runs using the DFS-4 code with
recovery being the inverse of the encoding (squares), 
average-case recovery over the whole $p$-range (triangles),
iterated average-case (diamonds; encoding optimized as well), and
iterated optimal-case recovery at each $p$ (circles). The last two are obtained by iterating
\refeq{gamopt}-\refeq{coptrlx} with the $\gam$-approximation
\refeq{gam approx} to 5 significant digits.
The results for the case when all 15 unitaries are randomly selected
are shown in Fig.~\ref{fig:5qu_ue_allrand}. Here we start from the
$[5,1,3]$ code and show standard recovery (squares) {\it vs} optimized
results (same legend as in Fig.~\ref{fig:5qu_ue_standard}).
The insets in Figs.~\ref{fig:5qu_ue_standard} and \ref{fig:5qu_ue_allrand} show the
fidelity gain relative to recovery via decoding in the DFS-4 case, or
relative to standard recovery ($[5,1,3]$ case). In the
iterated case, where the encoding is also optimized, the gain is
uniformly an order of magnitude or more. It is important to note that
this gain is obtained already for small values of $p$, where standard
encoding and recovery are designed to perform well. The encoding matrices in the
iterated case have support over all basis
states and and do not appear to yield previously known codes.

%%%%%%%%%%%%%%%%%%%%%%%%%%
{\it Discussion}.--- We have presented an optimization approach to
quantum error correction that yields codes which achieve robust
performance. One way to interpret the results presented in Figs.~1-3
is to consider a scenario where one is faced with a noise
channel with weight-2 or 3 errors, and can only use 4 or 5 qubits to
encode one. Without optimization
perhaps the most reasonable choices are the DFS-4 and [5,1,3] codes. However, our
results show that one can obtain
at least an order of magnitude higher fidelities by
optimizing both the encoding and recovery procedures, while respecting
the constraint of 4 or 5 qubits. There are of course standard codes dealing with
weight-2 and higher errors, but they require significantly more
qubits. We stress again that knowing that such performance is possible
can alleviate the need for unnecessary additional codespace which may
be impractical, and can help in improving the fault tolerance
threshold.

We note that the optimization approaches presented here
have differing computational costs.  Evaluating this cost depends on
the optimization algorithm and the problem structure
\cite{BoydV:04}. The cost can vary greatly if the algorithm is
modified for the specific problem structure. For general comparison 
purposes, all measures of computational complexity clearly will depend
on the dimensions of the various search spaces. This is summarized in
Table \ref{tab:optvars1} which gives the number of optimization
variables in the SDP optimizations. Clearly, the approximate indirect
method enjoys superior scaling.

An intriguing prospect is to integrate the results found
here within a complete ``black-box'' error correction scheme, that
takes quantum state or process tomography as input and iterates until
it finds an optimal error correcting encoding and recovery. Another
important open problem is to extend the robust procedures developed
here into a fault-tolerant error correction scheme.

\begin{table}[h]
\btab{|c|c||c|c|}
\hline
\multicolumn{2}{|c||}{Method}
&
Recovery
&
Encoding
\\
\hline\hline
Direct & Primal \refeq{favg opt}
& $(\ns^2-1)\nc^2$ & $(\nc^2-1)\ns^2$ 
\\ 
\cline{2-4} & Dual \cite{KosutL:06,Fletcher:06} & $\nc^2$ & $\ns^2$ 
\\ 
\hline\hline Indirect & $\gam$ opt. \refeq{gamopt} 
& $\me^2 = (\nca\nra)^2$ & $\ns\nc$ 
\\ 
& $\gam$ approx. \refeq{gam approx} & 0 & $\ns\nc$ 
\\ 
\hline 
\etab 
\caption{Number of optimization variables ($\nc=\ns\nca$).} 
\label{tab:optvars1} 
\end{table}

{\it Acknowledgments}.--- Funded under the DARPA QuIST Program and (to
D. A. L.) NSF CCF-0523675 and ARO W911NF-05-1-0440. 
We thank I. Walmsley, D. Browne, C. Brif, M. Grace, and H. Rabitz for
enlightening discussions.
%%

%%%%%%%%%%%%%%%%%%%%%%%%%%%%
%\bibliographystyle{prsty}
%\bibliography{rlk}
%\bibliography{D:/robert/tex/rlk}

%%%%%%%%%%%%%%%%%%%%%%%%%%%
\end{document}